\documentclass[aps,pra,twocolumn,superscriptaddress]{revtex4-2}
\usepackage[utf8]{inputenc}
\usepackage{graphicx}
\usepackage{amsmath}
\usepackage{natbib}
\usepackage{subcaption}
\usepackage{float}
\begin{document}

\title{Dynamics of Heisenberg XYZ spin Quantum Battery}
\author{Disha Verma}
\email{vermadisha785@gmail.com}
\affiliation{Department of Physics, National Institute of Technology, Tiruchirappalli 620015, India.}

\author{Indrajith VS}
\affiliation{Department of Physics, Mar Ivanios College, Thiruvananathapuram   695015, India.}

\author{R. Sankaranarayanan}
\affiliation{Department of Physics, National Institute of Technology, Tiruchirappalli 620015, India.}

\date{\today}

\begin{abstract}
   Spin systems have been extensively studied to understand the mechanisms of quantum batteries, which have shown the ability to charge faster than classical counterparts, even in closed systems. However, the internal dynamics of quantum batteries can significantly affect their performance, making it crucial to understand the influence of various parameters. In this study, we focus on the $XYZ$ Heisenberg spin system, examining key factors such as anisotropy in spin interactions and external magnetic field to optimize work output to ensure effective charging.
\end{abstract}

\maketitle

\maketitle
\section{Introduction}
Energy storage remains a key constraint in our technological landscape.  While conventional batteries have seen steady improvements, their fundamental limitations rooted in classical electrochemistry restrict their potential for the ever-growing demands of electric vehicles, renewable energy integration, and portable electronics.  Recently, the field of quantum thermodynamics has explored the possibility of leveraging the principles of quantum mechanics to revolutionize energy storage.  This leads to a new paradigm, known as quantum batteries ($QBs$), holds the promise of surpassing classical batteries in terms of efficiency, capacity, and even charging speed \cite{quach2023quantum}.
The concept of $QBs$ emerged in the last decade with theoretical proposals demonstrating the feasibility of storing energy in the quantum states of atoms and molecules \cite{alicki2013entanglement}.

These theoretical frameworks suggested intriguing advantages, such as ``superextensive charging", where larger batteries could charge faster than smaller ones, defying the limitations of classical systems.  The past decade have witnessed significant progress in the theoretical understanding of $QBs$, exploring diverse physical systems and optimizing charging protocols \cite{andolina2018charger,dou2022highly}. Pioneering research has explored into various aspects of quantum batteries \cite{binder2018thermodynamics}, such as the utilization of Dicke states \cite{dou2022extended}, the significance of entanglement in work extraction \cite{hovhannisyan2013entanglement,binder2015quantacell}, and nonlocal charging mechanisms for enhanced power storage \cite{gyhm2022quantum}. Furthermore, quantum batteries have made use of interacting spin systems \cite{ghosh2020enhancement}, which are charged through local magnetic fields. Importantly, these groundbreaking ideas have been effectively implemented in diverse platforms, including solid-state systems that feature individual two-level systems within single cavities or utilize ensembles of such systems  \cite{ferraro2018high}. An experimental study on XXZ Heisenberg quantum batteries shows coherence plays a crucial role in charging efficiency, challenging the emphasis on entanglement \cite{huang2023demonstration}. A paradigmatic model of a quantum battery has been experimentally implemented based on an organic microcavity \cite{quach2022superabsorption}.\\
The XYZ Heisenberg spin model has emerged as a powerful tool for investigating the theoretical basis of $QBs$.  This model captures the essential interactions between the battery's constituent parts and its environment, enabling researchers to analyze the impact of various factors on battery performance. Building upon previous work exploring the performance of Heisenberg XYZ spin chains \cite{ghosh2020enhancement}, this study investigates deeper into the role of external field and parameters involved in a two-spin system $QB$. Here we present an effective parameter for this model that characterizes the maximum energy storage (extractable work) within the battery.  This observation underscores the critical role of external field in manipulating the energy storage mechanism of $QBs$, highlighting the importance of optimizing the battery's Hamiltonian parameters.\\
The manuscript is organised as follows, section \ref{the} lays the groundwork by defining the system's energy (Hamiltonian). Section \ref{thee} analyzes charging: initial state  and energy stored through unitary evolution. Section \ref{uh} discusses the results, focusing on maximum energy storage.

\section{The model of battery}\label{the}
We consider the Hamiltonian of two spin  - $\frac{1}{2} $ system as $H = H_S + H_I$. Here the model for quantum battery is the Heisenberg XYZ system whose hamiltonian is defined as  
 \begin{equation}
 H_S = \frac{J}{2}[(1+\gamma)\sigma_{x}^{1}\sigma_{x}^{2}+(1-\gamma)\sigma_{y}^{1}\sigma_{y}^{2}]+  \frac{1}{2} J_{z}\sigma_{z}^{1}\sigma_{z}^{2} 
 \end{equation}\\
 where $\sigma_{k} $ are the Pauli spin matrices, $\gamma = (J_{x}-J_{y})/(J_{x}+J_{y})$ is the anisotropy in $XY$ plane. Here $J$ and $J_z$ are the strength of interaction in respective spin components. The interaction Hamiltonian is given by $H_I = \frac{1}{2}[B(\sigma_{z}^{1}+\sigma_{z}^{2})]$ where $B$ is  the strength of magnetic field. The matrix form of the Hamiltonian ($H$) in standard two qubit computational basis is given as
\begin{equation}
H=
\begin{pmatrix}
\frac{J_z}{2}+B & 0 & 0 & \gamma J\\
0 & -\frac{J_{z}}{2} & J & 0\\
0 & J & -\frac{J_{z}}{2} & 0\\
\gamma J & 0 & 0 & \frac{J_z}{2}-B
\end{pmatrix}.
\end{equation}
The above Hamiltonian is diagonalized and the corresponding eigenvalues  and eigenvectors are 
\begin{equation}
E_{1,2}=\frac{1}{2} (-J_{z}\pm 2J) ,\hspace{1cm}|\psi_{1,2}\rangle = \frac{1}{\sqrt{2}}\bigg(\pm\vert01\rangle + \vert10\rangle\bigg) \nonumber
\end{equation}
\begin{equation}
E_{3,4}=\frac{J_{z}}{2}\pm \eta , \hspace{1cm}\vert\psi_{1,2}\rangle=N_{\pm }\bigg(\frac{B\pm \eta}{\gamma J}\vert00\rangle+\vert11\rangle\bigg)
\end{equation}
where $\eta = \sqrt{B^2 + (\gamma J)^2}$, and the normalization constant $N_{\pm} = \left(({B \pm \eta}/{\gamma J})^2 + 1\right)^{-\frac{1}{2}}$. For $B = 0$ and $\gamma = 0,~ J = J_{z}$, the Hamiltonian corresponds to the Heisenberg spin with isotropic interaction, and the eigenfunctions are reduced to maximally entangled Bell states.

The thermal state or Gibbs state of this Hamiltonian is given by $\rho(T) = e^{- \beta H}/\mathcal{Z}$, and $\mathcal{Z} = \text{Tr} (e^{- \beta H})$ is the partition function where $\beta=1/{k_{B} T}$. It should be noted here that the energy of the system is scaled by setting $k_B T =1$ with $k_B$ being the Boltzmann constant, $T$ is the equilibrium temperature. The thermal state in the computational basis is obtained as \cite{indrajith2019entanglement}:

\begin{equation}
 \rho(T) = \frac{1}{\mathcal{Z}}
 \begin{pmatrix}
 \mu_{-} & 0 & 0 & \kappa \\
 0 & \nu & \epsilon & 0 \\
 0 & \epsilon & \nu & 0 \\
 \kappa & 0 & 0 & \mu_{+}
 \end{pmatrix}
 \label{Thermal}
\end{equation}

with matrix elements $\mu_{\pm}=e^{-{J_{z}}/{2}}(\cosh \eta ~{\pm}~{B}/{\eta}\sinh\eta)$, $\kappa=-{\gamma J}/{\eta}e^{{-J_{z}}/{2}}\sinh\eta$, $\nu=e^{{J_{z}}/{2}}(\cosh J )$, $\epsilon=-e^{{J_{z}}/{2}}\sinh J$ and the partition function $\mathcal{Z}=2(e^{-{J_{z}}/{2}}\cosh\eta+e^{{J_{z}}/{2}}\cosh J)$. 
 
 \section{Energy stored in Quantum Charging}\label{thee}
\subsection{Initial Battery State}\label{ther}

In the initial preparation phase, the quantum battery can be in one of two distinct states: the ground state or the Gibbs state. In the first scenario, the battery assumes the ground state of the normalized Hamiltonian, representing a state corresponding to absolute zero temperature. This configuration sets the foundation for understanding the quantum properties of the battery under no thermal influence.

On the other hand, in the second scenario the battery is prepared in Gibbs state (also called as canonical equilibrium state) characterized by a finite temperature $T$. In this state, the battery experiences thermal effects, and its properties are influenced by the thermal distribution of energy levels. The partition function $\mathcal{Z}$ encapsulates the statistical sum of all quantum states weighted by the corresponding Boltzmann factors. This canonical equilibrium state provides valuable insights into the quantum battery's behavior under realistic thermal conditions, offering a more comprehensive understanding of its performance characteristics. The dichotomy between the ground state and the canonical equilibrium state serves as a foundational consideration in the study of quantum batteries, enabling exploration of their quantum features at zero and finite temperature regimes. It has been observed \cite{ghosh2020enhancement} that tuning system parameters could lead to maximal power generation from a state that is initially prepared at finite temperature than the state with absolute zero temperature.

\subsection{Unitary charging and energy stored }\label{yh}
In the charging process of the quantum battery within a closed system, the evolution of state is dictated by the unitary operation. Besides reducing the amount of heat generated during the charging process, the implementation of unitary charging techniques has been shown to induce non-classical correlations within many-body quantum batteries \cite{campaioli2024colloquium}. This results in  significant enhancement of the charging power scaling, surpassing the conventional limits typically encountered in classical battery systems. The amount of energy that can be stored and extracted can be calculated from the time evolution  \begin{equation}
    \dot{\rho}(t) = -\frac{i}{\hbar}[H + H_c(t),\rho(t)] \label{eq:h}
\end{equation} 
where $H_c$ is hermitian time dependent charging that is turned on at time $t=0$ and off at time $t$ and $\hbar$ is the Planck's constant. The energy ($W$) stored in such way, reaching some arbitrary state $\rho$($t$) from some initial state $\rho$($0$) is measured with respect to battery Hamiltonian $H$ is
\begin{equation}
   W(t) = \text{Tr}[H \rho(t)] - \text{Tr}[H \rho(0)]  \label{eq:j}
\end{equation}
where $\rho(t) = U(t;0) \rho(0) U^\dagger(t;0)$ is the time evolved state obtained from the solution of eq.(\ref{eq:h}). Here $ U(t;0) = \mathcal{T} exp{\left\{ -\frac{i}{\hbar} \int_{0}^{t} ds \, [H_0 + H_c(s)] \right\}}$  is the time-evolution operator where $\mathcal{T}$ is the time-ordering operator. In what follows, we set $\hbar=1$.  It is important to consider that when focusing solely on unitary evolution, the processes of work injection (charging) and extraction can be seen as essentially identical tasks \cite{campaioli2024colloquium}. The maximum amount of work that can be extracted from a quantum battery is given by ergotropy \cite{allahverdyan2004maximal}, which is same as $W(t)$ if the initial state $\rho(0)$ is the lowest possible energy state (passive state).  It is because for the entropy-preserving unitary evolution process the maximum amount work that can be extracted is equal to the energy stored in the system \cite{bera2019thermodynamics,konar2022quantum}. In general, the stored energy does not coincide with the ergotropy for a battery Hamiltonian in contact with the environment \cite{quach2020using,farina2019charger},  where the evolution is nonunitary. 

In this quantum battery model the fundamental framework involves the time evolution of the initial state using the time independent charging Hamiltonian $H_c$ over a total time $t$. The specific charging Hamiltonian used in this study is $H_c = \omega S_x$, represents a constant magnetic field applied along the $x$ axis, where $\omega$ is the strength of applied field, and $S_x = \frac{1}{2} (\sigma_x^{1} + \sigma_x^{2})$ is the  total spin operator along $x$ axis.

The unitary transformation in case of a time independent charging Hamiltonian, $U = e^{-iH_c t}$ governs the evolution of the quantum state during the charging period. This transformation, applied to the initial state $\rho(0)=\rho(T)$ is resulting to the state $\rho(t)$ at later time $t$. Since $\rho(T)$ is a unique passive state of the Hamiltonian $H$, $W(t)$ is the limiting value of maximum extractable work \cite{alicki2013entanglement}.
\section{Results and Discussion}\label{uh}
\subsection{Charging Dynamics }
We begin our discussion by obtaining an analytical expression for the energy stored as
\begin{equation}
  W (t) =(a+b) -b \cos{2 \omega t}-a \cos{\omega t}   \label{l}
\end{equation}
where 
\[a = \frac{4 B^2 \sinh{\eta}}{d}\]
\[b = \frac{b_1 \cosh{J} + b_2 \cosh{\eta}}{d} + \frac{b_3 (-e^{J_z} \eta \sinh{J} + J \gamma \sinh{\eta})}{d}
\]with  $b_1 = e^{J_z} (J_z + J(1+\gamma)) \eta $,   $b_2 = (-J_z + J(1+\gamma)) \eta$,  $b_3 = J(-1+\gamma)$ and $d= \eta(e^{J_z} \cosh{J}+\cosh{\eta})$.
Thus the energy stored in the system (battery) is periodic with frequency $\omega$,  as we intuitively expect.

\begin{figure}[H]
    \centering
    \includegraphics[width=0.45\textwidth]{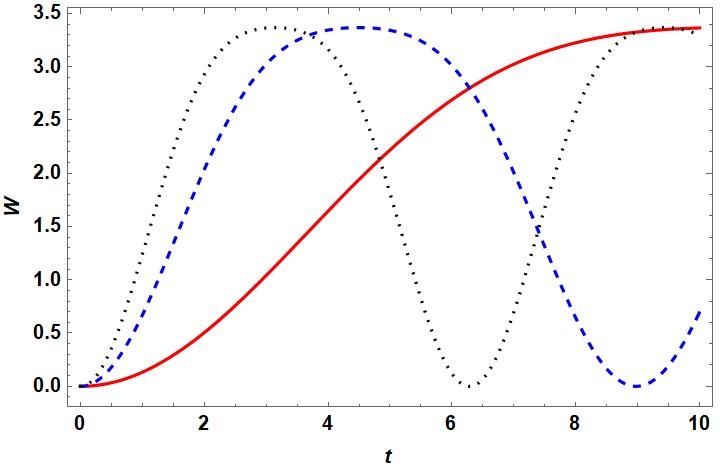}
    \caption{The specific parameter values chosen are  $B=1$, $J_z=0.2$, $J=0.2$, $\gamma=0.5$, $\omega=1$, $0.7$, and $0.3$ for the black (dotted), blue (dashed), and red (solid) respectively.}
    \label{fig:grap1}
\end{figure}

Generally, reducing the strength of the field can lead to a more gradual process of charging and energy storage. This is primarily because the external field has an impact on the energy levels of the quantum system, and a weaker field tends to cause smaller shifts in energy levels during the charging cycles. The slowed-down process of energy storage may present various benefits for quantum batteries. Notably, it can enable a more precise regulation of energy transfer, thus decreasing the chances of energy loss or degradation. In   Fig.~\ref{fig:grap1}, we depict the oscillation of energy stored over time for specific chosen parameter values. The graph exhibits a periodic pattern with a time period of $\Tilde{t}= 2\pi/\omega$, at which $\rho(\Tilde{t})=\rho(0)$.

This result implies that achieving a greater amount of energy stored in a shorter time frame necessitates a stronger external magnetic field. However, a slow and stable charging can be realised with weaker magnetic field. This enhancement contributes to increased energy efficiency and a prolonged lifespan for the battery. Additionally, a slower pace of charging  provides the quantum system with more time to sustain coherence, a crucial characteristic for efficient energy storage and transfer in quantum batteries. Coherence, in this context, pertains to the preservation of phase relationships among different elements of the quantum system. A coherent system is better equipped to store and release energy effectively \cite{kamin2020entanglement}.\\  

\subsection{Maximum work }

From the simple form of $W(t)$ given by eq.(\ref{l}), the maximum energy stored can be calculated as  \[W_{max1} = 2a    \quad,\;   \text{if}  \left| \frac{a}{4b} \right| > 1 \label{j1
}\]
\begin{equation}
W_{max2} = 2b + a + \frac{a^2}{8b} \quad,\;\text{if}  \left| \frac{a}{4b} \right| \leq 1. \label{j2}
\end{equation}
\begin{figure}[H]
\centering

\includegraphics[width=0.5\textwidth]{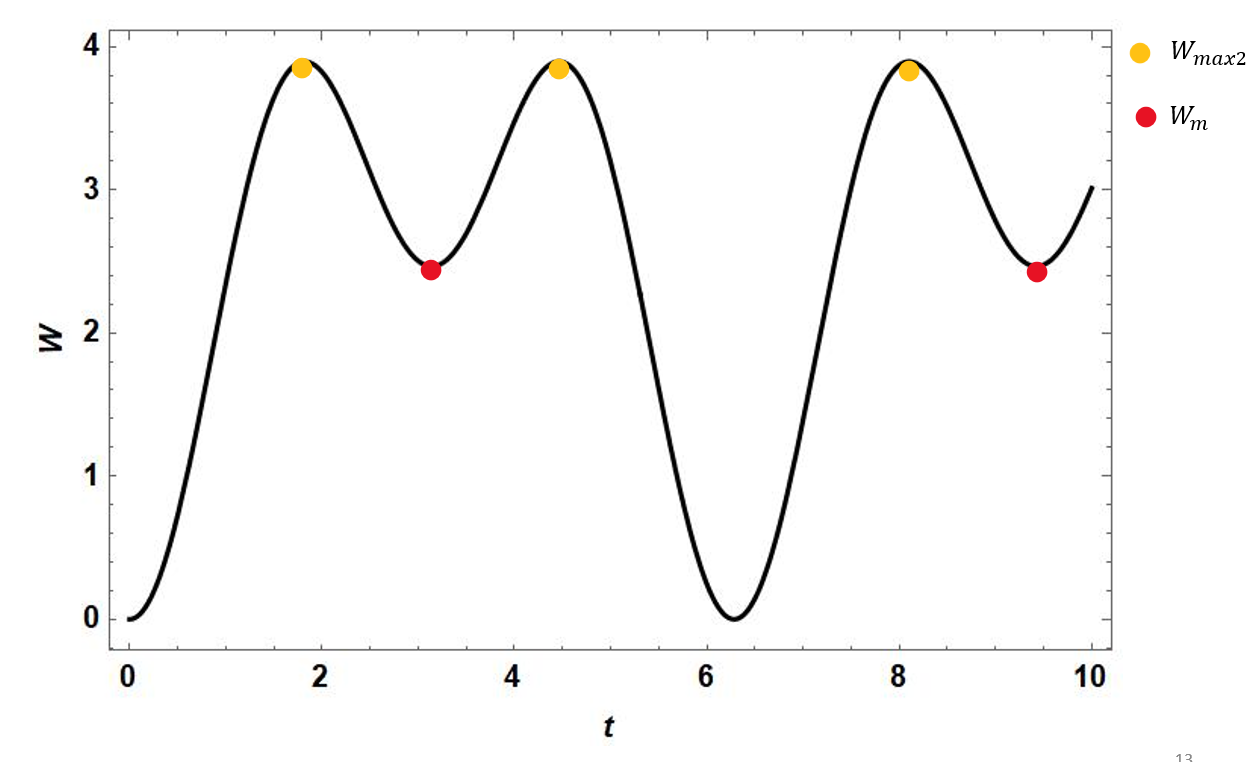}
\caption{The specific parameter values chosen are such that $\left|{a}/{4b}\right| \leq 1$, so the maximum work is $W_{\text{max2}}$. The first maximum occurs at $t_2 = ({1}/{\omega})\cos^{-1}\left({-a}/{4b}\right)$, and second maximum occurs at $t_2 = 2\pi - ({1}/{\omega})\cos^{-1}\left(-{a}/{4b}\right)$, respectively within one oscillation. $W_m$ here corresponds to the minimum observed within one period.}
  \label{fig:graph9}

\end{figure}
The corresponding time for the two different cases are 
\( t_1 = n_1 \pi/\omega \), where \( n_1 \) is an odd integer, and 
\( t_2 = 2n_2 \pi \pm (1/\omega) \cos^{-1}\left( -a/4b) \right) \), where \( n_2 \) is an integer.
We shall note that Fig. \ref{fig:grap1} correspond to the case of $\left|{a}/{4b}\right| > 1$ and $ W_{max1}$ = 3.369, with $t_1$ = {$n_1 \pi$}/{$\omega$}. On the other hand, the energy stored for the case $\left|{a}/{4b}\right| \leq 1$ (Fig. \ref{fig:graph9}) shows two maxima separated by a time interval $\Delta t = 2 \pi - ({2}/{\omega}) \cos^{-1}\left({-a}/{4b}\right)$ within one time period of the cycle. We also observe that the minimum $W_m=2a$ is same as $W_{max1}$. This minimum in work becomes $W_{max1}$ as $B$ increases such that $\left|{a}/{4b}\right| > 1$. Interestingly, it is observed that as the factor ${a}/{4b}$ $\to 1$, the difference  $W_{max2}-W_{m} = 1/2 +(1/2)(a/4b)^2 - a/4b $ $\to$ $0$ such that  $\Delta t$ $\to$ $ 2\pi (1-1/ \omega)$. In other words, maximum extractable work is fairly stable over a finite time $\Delta t$ whose limiting value is $2 \pi$ as $\omega \to \infty$. This feature is shown in Fig.~\ref{fig:fig9} at $B=1$, for which ${a}/{4b}$ $\approx 0.9$.

\begin{figure}[H]
    \includegraphics[width=0.5\textwidth]{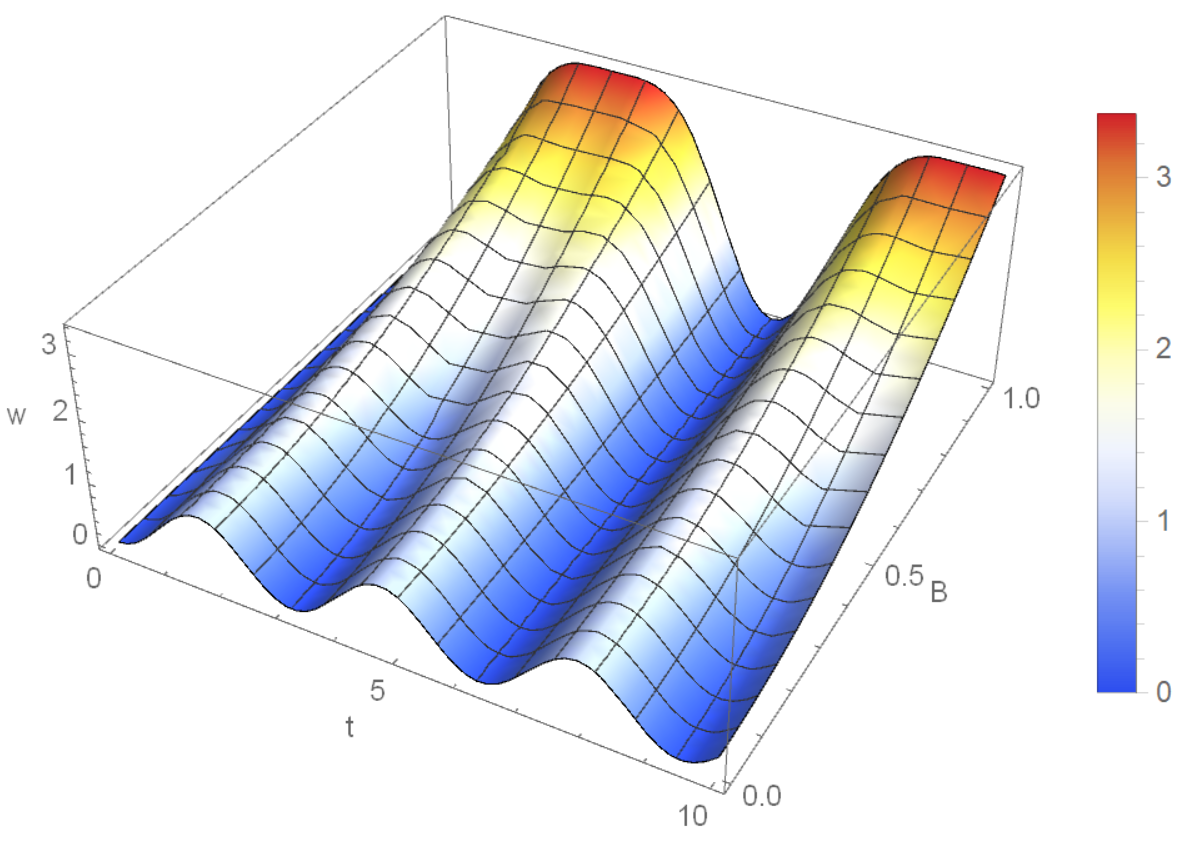}
     \caption{Plot shows the evolution of stored energy for different magnetic field $B$.}
    \label{fig:fig9}
\end{figure}

\begin{figure}[H]
    
\begin{minipage}{0.45\textwidth}
        \centering
        \includegraphics[width=\textwidth]{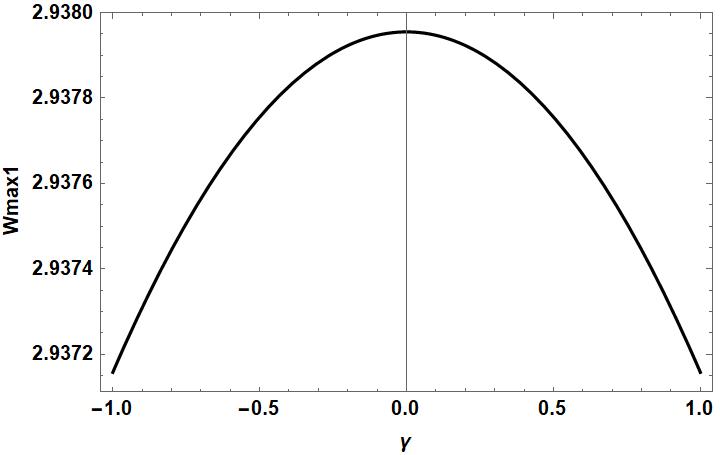}
        \textbf{(a)} 
    \end{minipage}\hfill
    \begin{minipage}{0.45\textwidth}
        \centering
        \includegraphics[width=\textwidth]{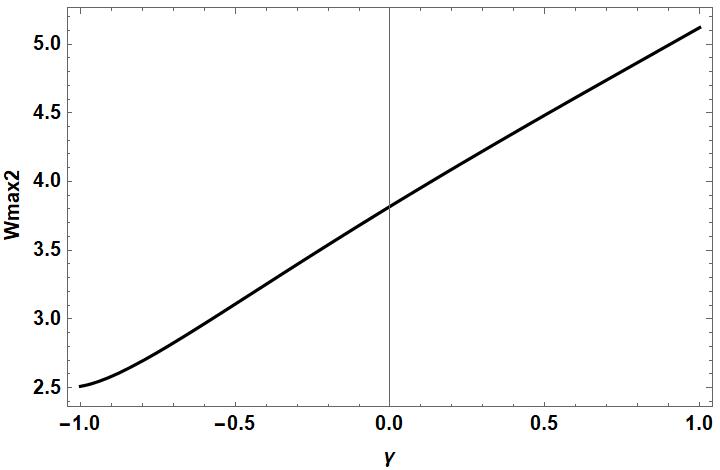}
        \textbf{(b)} 
    \end{minipage}
   
    \caption{(a) and (b) shows the dependence of  maximum energy stored in the system on anisotropy parameter $\gamma$.}
    \label{fig:stable}
\end{figure}

\begin{figure}[H]
    
\includegraphics[width=0.5\textwidth]{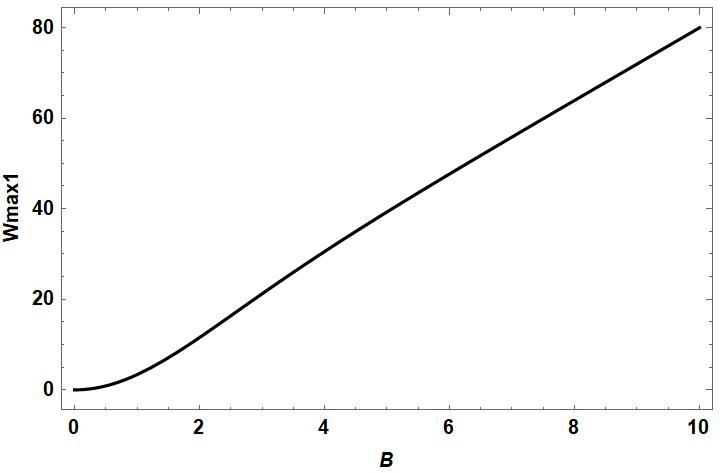}
     \caption{Plot shows the maximum stored energy for higher values of magnetic field $B$.}
    \label{fig:fig7}
\end{figure}

Fig. \ref{fig:stable}(a)  shows the dependence of $W_{max1}$ on the anisotropy parameter $\gamma$. It should be noted that  Fig. \ref{fig:stable}(b)  shows the behavior for  low values of $B$, such that as $B$ increases (keeping all other parameters constant) Fig. \ref{fig:stable}(a) becomes relevant since the condition $\left|{a}/{4b}\right| > 1$ is fulfilled. It is clear from Fig. \ref{fig:stable}(a) and \ref{fig:stable}(b) that while the dependence of $\gamma$ on $W_{max1}$ is insignificant, $W_{max2}$ varies significantly with $\gamma$. In other words, the role of anisotropy becomes important as $B$ is decreased.  Also it is clear that the parameter $B$ in the system holds a crucial role in determining the maximum stored work in the battery, as depicted in Figure \ref{fig:fig7}, and the same can be understood as follows.
Recollecting that 
\[ 
W_{max1}= \frac{8 B^2 \sinh \eta}{d} 
\]
the $B$ dependence for large $B$ is more evident with the following approximation:
Since $\eta$ $\approx$ $B$ for large $B$,
\[ W_{max1} \approx \frac{8 B^2 \sinh B}{B (e^{J_z} \cosh J + \cosh B)} \, .\]
Further, for large $B >0$, \(\cosh B \approx \sinh B\) and hence,
\begin{equation}
    W_{max1} \approx \frac{8 B^2 \sinh B}{B \sinh B}  = 8B. \label{j3}
\end{equation}
Therefore, \( W_{max1} \) is linearly dependent on \( B \) for the large values of $B$. This emphasizes the role of $B$ on the maximum energy stored in two-qubit XYZ spin as quantum battery. 

\section{Conclusion}\label{con}
The manipulation of external field in Heisenberg XYZ spin quantum battery plays a crucial role in influencing energy storage dynamics. Our study reveals that when the battery is coupled with a constant external field which acts as a charging, the energy stored within it exhibits oscillatory behavior. Through detailed analysis, we have identified two distinct oscillations of energy storage
in terms of an effective parameter of the battery system. Here we have demonstrated that one of the oscillations can be tuned for a maximum stable energy storage or work extraction over a finite time duration. It is also observed that the dependence of maximum work on anisotropy
is significant only when the magnetic field of the battery is low. Finally, it is shown that the maximum extractable work is eight times the strength of the magnetic field of the battery. In conclusion, this study reveals the significance of the internal parameters of the XYZ spin system
from the quantum battery perspective.

\bibliographystyle{unsrt}
\bibliography{ref}

\begin{thebibliography}{10}

\bibitem{quach2023quantum}
James~Q Quach, Giulio Cerullo, and Tersilla Virgili.
\newblock Quantum batteries: The future of energy storage?
\newblock {\em Joule}, 7(10):2195--2200, 2023.

\bibitem{alicki2013entanglement}
Robert Alicki and Mark Fannes.
\newblock Entanglement boost for extractable work from ensembles of quantum
  batteries.
\newblock {\em Physical Review E}, 87(4):042123, 2013.

\bibitem{andolina2018charger}
Gian~Marcello Andolina, Donato Farina, Andrea Mari, Vittorio Pellegrini,
  Vittorio Giovannetti, and Marco Polini.
\newblock Charger-mediated energy transfer in exactly solvable models for
  quantum batteries.
\newblock {\em Physical Review B}, 98(20):205423, 2018.

\bibitem{dou2022highly}
Fu-Quan Dou, Yuan-Jin Wang, and Jian-An Sun.
\newblock Highly efficient charging and discharging of three-level quantum
  batteries through shortcuts to adiabaticity.
\newblock {\em Frontiers of Physics}, 17:1--9, 2022.

\bibitem{binder2018thermodynamics}
Felix Binder, Luis~A Correa, Christian Gogolin, Janet Anders, and Gerardo
  Adesso.
\newblock Thermodynamics in the quantum regime.
\newblock {\em Fundamental Theories of Physics}, 195:1--2, 2018.

\bibitem{dou2022extended}
Fu-Quan Dou, You-Qi Lu, Yuan-Jin Wang, and Jian-An Sun.
\newblock Extended dicke quantum battery with interatomic interactions and
  driving field.
\newblock {\em Physical Review B}, 105(11):115405, 2022.

\bibitem{hovhannisyan2013entanglement}
Karen~V Hovhannisyan, Mart{\'\i} Perarnau-Llobet, Marcus Huber, and Antonio
  Ac{\'\i}n.
\newblock Entanglement generation is not necessary for optimal work extraction.
\newblock {\em Physical review letters}, 111(24):240401, 2013.

\bibitem{binder2015quantacell}
Felix~C Binder, Sai Vinjanampathy, Kavan Modi, and John Goold.
\newblock Quantacell: powerful charging of quantum batteries.
\newblock {\em New Journal of Physics}, 17(7):075015, 2015.

\bibitem{gyhm2022quantum}
Ju-Yeon Gyhm, Dominik {\v{S}}afr{\'a}nek, and Dario Rosa.
\newblock Quantum charging advantage cannot be extensive without global
  operations.
\newblock {\em Physical Review Letters}, 128(14):140501, 2022.

\bibitem{ghosh2020enhancement}
Srijon Ghosh, Titas Chanda, Aditi Sen, et~al.
\newblock Enhancement in the performance of a quantum battery by ordered and
  disordered interactions.
\newblock {\em Physical Review A}, 101(3):032115, 2020.

\bibitem{ferraro2018high}
Dario Ferraro, Michele Campisi, Gian~Marcello Andolina, Vittorio Pellegrini,
  and Marco Polini.
\newblock High-power collective charging of a solid-state quantum battery.
\newblock {\em Physical review letters}, 120(11):117702, 2018.

\bibitem{huang2023demonstration}
Xiaojian Huang, Kunkun Wang, Lei Xiao, Lei Gao, Haiqing Lin, and Peng Xue.
\newblock Demonstration of the charging progress of quantum batteries.
\newblock {\em Physical Review A}, 107(3):L030201, 2023.

\bibitem{quach2022superabsorption}
James~Q Quach, Kirsty~E McGhee, Lucia Ganzer, Dominic~M Rouse, Brendon~W
  Lovett, Erik~M Gauger, Jonathan Keeling, Giulio Cerullo, David~G Lidzey, and
  Tersilla Virgili.
\newblock Superabsorption in an organic microcavity: Toward a quantum battery.
\newblock {\em Science advances}, 8(2):eabk3160, 2022.

\bibitem{indrajith2019entanglement}
VS~Indrajith, R~Muthuganesan, and R~Sankaranarayanan.
\newblock Entanglement and measurement-induced quantum correlation in
  heisenberg spin models.
\newblock {\em Physica A: Statistical Mechanics and its Applications},
  527:121325, 2019.

\bibitem{campaioli2024colloquium}
Francesco Campaioli, Stefano Gherardini, James~Q Quach, Marco Polini, and
  Gian~Marcello Andolina.
\newblock Colloquium: quantum batteries.
\newblock {\em Reviews of Modern Physics}, 96(3):031001, 2024.

\bibitem{allahverdyan2004maximal}
Armen~E Allahverdyan, Roger Balian, and Th~M Nieuwenhuizen.
\newblock Maximal work extraction from finite quantum systems.
\newblock {\em Europhysics Letters}, 67(4):565, 2004.

\bibitem{bera2019thermodynamics}
Manabendra~Nath Bera, Arnau Riera, Maciej Lewenstein, Zahra~Baghali Khanian,
  and Andreas Winter.
\newblock Thermodynamics as a consequence of information conservation.
\newblock {\em Quantum}, 3:121, 2019.

\bibitem{konar2022quantum}
Tanoy~Kanti Konar, Leela Ganesh~Chandra Lakkaraju, Srijon Ghosh, and Aditi Sen.
\newblock Quantum battery with ultracold atoms: Bosons versus fermions.
\newblock {\em Physical Review A}, 106(2):022618, 2022.

\bibitem{quach2020using}
James~Q Quach and William~J Munro.
\newblock Using dark states to charge and stabilize open quantum batteries.
\newblock {\em Physical Review Applied}, 14(2):024092, 2020.

\bibitem{farina2019charger}
Donato Farina, Gian~Marcello Andolina, Andrea Mari, Marco Polini, and Vittorio
  Giovannetti.
\newblock Charger-mediated energy transfer for quantum batteries: An
  open-system approach.
\newblock {\em Physical Review B}, 99(3):035421, 2019.

\bibitem{kamin2020entanglement}
FH~Kamin, FT~Tabesh, S~Salimi, and Alan~C Santos.
\newblock Entanglement, coherence, and charging process of quantum batteries.
\newblock {\em Physical Review E}, 102(5):052109, 2020.

\end{thebibliography}

\end{document}